\documentstyle[12pt,twoside,fleqn,espcrc1]{article}
\title{Darwin-Foldy term and proton charge radius}  

\author{M. Bawin\address{Universit\'{e} de Li\`{e}ge, Institut de
Physique B5, Sart Tilman, 4000 Li\`{e}ge 1, Belgium}
and S. A. Coon\address{Physics Department, New Mexico State University,
         Las Cruces, NM  88003, USA}}

\begin{document}

\maketitle

 Back in 1958 Foldy~\cite{Foldy}, following Pauli's early work on
incorporating the anomalous magnetic moment of a particle into the
Dirac equation, proposed a systematic way
 of including finite size
effects in the description of a
 relativistic spin
$\textstyle\frac{1}{2}$ particle in interaction with an
 external
electromagnetic field by means of the Dirac equation. An
 apparently
perturbing feature of Foldy's treatment when applied to a
 neutron
is that the  magnitude of the neutron charge radius is
 accounted
for almost completely by treating the neutron as a {\it
 point}
Dirac particle (the ``Darwin-Foldy''term), thus apparently leaving
no
 room for finite size effects  \cite{RG} but rather a so-called
``intrinsic radius'' much smaller than the Compton wavelength of the
neutron.

We recently suggested that the composite nature of the neutron can be
described by including Dirac-Pauli form factors in the Dirac equation.
Then, the Darwin-Foldy term is cancelled exactly by a term in a
low-momentum expansion of the form factors, leaving the quantity
measured in neutron-electron scattering to be directly identified
with the neutron charge radius \cite{bc}. Our result agrees with
recent results by Isgur \cite{Isgur} (in the context of the valence
quark
 model) and Cardarelli and Simula \cite{CS_99} (in the
relativistic constituent
 quark model). Note however that the
interpretation of the Darwin-Foldy
 term for a particle with
structure remains a controversial topic
\cite{Byrne,Alex,Bun}. For example, one can read  in \cite{Byrne}
that the intrinsic radius ``provides a measure of the failure of the
Dirac-Pauli equation to account in detail for the observed properties
of the neutron at low energy''.

	In this contribution we study the Dirac equation for a
finite
 size proton in an external electric field with explicit
introduction
 of Dirac-Pauli form factors. Our aim is twofold. On
the one hand, we wish to study whether our conclusions regarding the
exact cancellation between Dirac form factor and  Foldy term
contributions occurring for the neutron still hold for the proton. On
the other hand, we wish to clearly illustrate  some of the specific
features of the description of a composite particle like the proton
with the Dirac equation. Indeed, the Dirac equation includes the effect
of virtual particle-antiparticle creation pair terms (Foldy's ``dancing
motion'') which endow even a charged {\em point} particle with an
effective size, the Darwin term. (The ``Darwin-Foldy'' term of the
neutron, however, arises from the large anomalous magnetic moment of
this neutral particle). Whether such an effect should be included in
the definition of the particle charge radius is at the heart of the
controversy surrounding the Dirac approach description of the nucleon
charge radius. 

Our main results are the following:\\ 1) The Dirac form
factor contribution continues to  cancel  the Darwin-Foldy 
contribution from the proton's {\em anomalous} magnetic moment  (as was
also demonstrated in the model of
 Ref.~\cite{CS_99}).  There
remains, however,
 the standard Darwin term (the $\frac{e}{8m^2}$
term) coming
 from the {\it normal} part of the proton magnetic
moment, a
 consequence of the {\em Zitterbewegung} of the charged
Dirac particle.\\ 2) The above cancellation crucially depends upon
using the Dirac equation with form factors to describe the finite size
and anomalous magnetic moment of the baryons. The resulting charge
radius is then that of the Sach's form factor \cite{bc} and is thus the
same as the definitions used in recent chiral perturbation theory
calculations \cite{UM}.

 Our starting point is the Dirac Hamiltonian H  for a finite size
proton in an external field with explicit introduction of Dirac-Pauli
form factors :      \begin{equation} H = \mbox{\boldmath $\alpha$}
\cdot {\bf p} + \beta
 m + i \frac{eF_2}{2m}\beta\mbox{\boldmath
$\alpha$} \cdot{\bf E} +
 eF_1V, \end{equation} where  $F_1$ and
$F_2$ are, respectively, the
 proton Dirac and Pauli form factors,
and  $V$ is the  potential
 associated with the external field ${\bf
E}$ in the Coulomb gauge.
 The connection between $F_1$, $F_2$, and
the electric and magnetic  Sachs form factors $G_E$ and $G_M$ is:
\begin{eqnarray}
	G_E(t) &=& F_1(t) + \frac{t}{4m^2} F_2(t) \nonumber \\
	G_M(t) &=& F_1(t) +F_2(t)  \label{eq:Sachs}
\end{eqnarray}
 The  Sachs form factors have  simple interpretations as the spatial
Fourier transforms of the nucleon's charge and magnetization
distributions in the Breit frame\cite{Sachs}, where $t = -{\bf q}\,^2$,
and $\bf q$ is the usual 3-dimensional momentum transfer. Specifically,
we have:
\begin{equation}
	\rho_P({\bf r}) = \left (\frac{1}{2\pi}\right)^3
	\int d^3 {\bf q}\, e^{+i {\bf q}\cdot { \bf r}}
		G_E(-{\bf q}^2)\;,  \label{eq:rhoGE}
\end{equation}		
such that the normalization $\int  d^3 {\bf r}\,\rho_P({\bf r})$ is one
for the proton. One defines the square of the proton charge radius
$r^2_{Ep}$ by the relation:  
\begin{equation}
 G_E(-{\bf q}^2) = 1 - \frac{1}{6} r^2_{Ep}{\bf q}^2  + \cdots\, .  \label{rp}
\end{equation}

Taking into account the relations \cite{bc}: 
\begin{equation} F_1(0)= 1,
\end{equation} 
\begin{equation} F'_1(0) \equiv
\frac{\partial{F_1({\bf q}^2=0)}}{\partial{{\bf q}^2}} = -\frac{1}{6}
r^2_{Ep} + \frac{\kappa}{4m^2},  \label{rsquare} 
\end{equation}
where $\kappa$ is the proton anomalous dimensionless
 magnetic
moment,  we eventually get the
 following equation from the
Foldy-Wouthuysen reduction of the Dirac
 Hamiltonian (1) together
with a low ${\bf q}\,^2$ expansion of $\tilde{V}({\bf q}) =  e 
F_1(-{\bf q}\,^2)V({\bf q})$ \cite{bc}:
 \begin{eqnarray} \{ \frac{{\bf p}^2}{2m} -
\frac{{e\bf p}}{2m^2}\cdot ({\bf E} \times \mbox{\boldmath $\kappa$}
) + \frac{e^2 \kappa^2 {\bf E}^2}{8m^3} - e(\frac{r^2_{Ep }}{6}+
\frac{1}{8m^2}) (\mbox{\boldmath $\nabla$}\cdot  {\bf E}) & &
\nonumber \\ + V + \frac{\mbox{\boldmath $\nabla$}^2}{8m^2} \tilde{V}
\} \psi & = E\psi\ ,  \label{FW}       
 \end{eqnarray}
 where $ \tilde{V}$ is given by: 
\begin{equation}
	\tilde{V}({\bf r}) \equiv eF_1V = e \left (\frac{1}{2\pi}\right)^3
	\int  d^3 {\bf q}\, e^{i {\bf q}\cdot
	{\bf r}} F_1
	(-{\bf q}\,^2)V({\bf q})\,.    \label{eq:efv}
\end{equation}         
In equation (\ref{FW}), the Dirac form factor contribution continues to
 cancel  the Darwin-Foldy  contribution from the proton's
anomalous
 magnetic moment  (as was also demonstrated in the model
of
 Ref.~\cite{CS_99}).  There remains, however, the standard
Darwin term (the $\frac{e}{8m^2}$ term) coming from the {\it
normal} part of the proton magnetic moment, a consequence of the
dancing motion   of the {\em charged} Dirac particle. By convention,
that term,  also  present for a structureless electron,
does not contribute to the charge radius. Thus the only structure term
in the coefficient of the external field charge density is the physical
charge radius square $r_{Ep}$. We therefore conclude that the inclusion
of the nucleon form factor together with the definition of the nucleon
charge radius given by equation (\ref{rp}) removes the apparent difficulties 
associated with using the Dirac equation to describe the nucleon charge
radius. Contrary to previous statements in the literature
(\cite{Wilets}), the Dirac equation does not necessarily fail at the
level of the Darwin-Foldy term.

\end{document}